The 9th International Conference on Ambient Systems, Networks and Technologies (ANT 2018)

# Enhancing Middleware-based IoT Applications through Run-Time Pluggable QoS Management Mechanisms. Application to a oneM2M compliant IoT Middleware

Clovis Anicet Ouedraogo[a], Samir Medjiah[a,b], Christophe Chassot[a,c,*], Khalil Drira[a]

[a] *CNRS, LAAS, 7 avenue du Colonel Roche, F-31400 Toulouse, France*
*Univ. Toulouse, [b] UPS, [c] INSA, LAAS, F-31400 Toulouse, France*
*{ouedraogo, medjiah, chassot, drira} @ laas.fr*

**Abstract**

In the recent years, telecom and computer networks have witnessed new concepts and technologies through Network Function Virtualization (NFV) and Software-Defined Networking (SDN). SDN, which allows applications to have a control over the network, and NFV, which allows deploying network functions in virtualized environments, are two paradigms that are increasingly used for the Internet of Things (IoT). This Internet (IoT) brings the promise to interconnect billions of devices in the next few years rises several scientific challenges in particular those of the satisfaction of the quality of service (QoS) required by the IoT applications. In order to address this problem, we have identified two bottlenecks with respect to the QoS: the traversed networks and the intermediate entities that allows the application to interact with the IoT devices. In this paper, we first present an innovative vision of a "network function" with respect to their deployment and runtime environment. Then, we describe our general approach of a solution that consists in the dynamic, autonomous, and seamless deployment of QoS management mechanisms. We also describe the requirements for the implementation of such approach. Finally, we present a redirection mechanism, implemented as a network function, allowing the seamless control of the data path of a given middleware traffic. This mechanism is assessed through a use case related to vehicular transportation.

## 1. Introduction

*The Internet of Things, its application and their QoS requirements*. The future Internet will include not only usual terminals but more generally any form of connected *objects* (or *things*) authorizing the development of new business activities, in various domains such as remote supervision, personal assistance, or urban transport. This IoT will also have to meet non-functional needs (e.g. quality of service - QoS, security) of these new applications.

* Corresponding author. Tel.: +33-561-337-816; fax: +33-561-559-500.
 *E-mail address*: chassot@laas.fr

The interactions between the underlying application software(s) and the connected objects will be based on heterogeneous networks and on *middleware* layers. Indeed, from 2010, a major standardization effort has been conducted, notably via the ETSI and the oneM2M consortium[1,2]. The resulting frameworks are aimed at abstracting applications from complexity of the underlying technologies (networks and objects); they are also aimed at avoiding *vertical* fragmentation of currently developed IoT solutions thanks to a generic middleware layer. Based on the REST architectural style (http, CoAP, …), this framework makes them appear what can be called *Middleware (MW) nodes*, named *gateway* and *server* in the ETSI vision, and *mn-cse* and *in-cse* in the oneM2M vision (Fig. 1). Both visions make also appear two major bottlenecks with regard to QoS considerations (service availability, bounded response time, etc.): within the connected objects and IP networks, and within the MW nodes.

In this context, several attempts have already been done to face (among other) the QoS requirements at the middleware level. Those attempts are based, for instance, on the deployment of QoS-oriented mechanisms within the MW nodes, with the aim to manage the application traffic (e.g. delaying less priority http requests in case of congestion) and / or allocated resources of the underlying machines[3]. Those propositions have also shown the benefits that can be induced by a deployment of such mechanisms "outside" of the data path, for instance thanks to intermediate proxies configured (for instance) as traffic load balancer / shaper / dropper (Fig. 2).

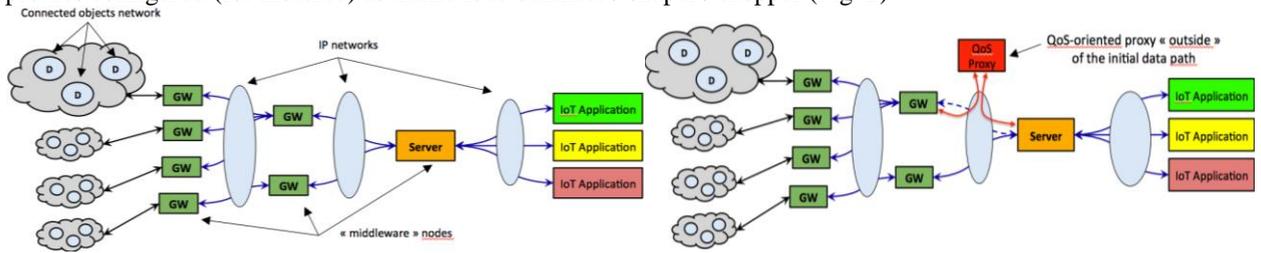

Fig. 1: Overview of an IoT Middleware    Fig. 2: QoS-oriented middleware with proxy outside of the data path

*Network function virtualization.* Since a few years, *Cloud* and now *Fog Computing* constitute opportune environments to help meeting IoT applications' functional needs. More generally, the advent of *virtualization* technologies makes it now possible the deployment of (e.g. QoS-oriented) mechanisms on dedicated equipment but also on private or public *data centers* having hypervisors offering the required functional capabilities. The concept of *virtual network function* (VNF) has been defined by the ETSI as part of its work on standardization of the NFV technology[4], the term "virtual" meaning that a NF is not necessarily implemented on a dedicated equipment. This concept is today to be considered in a wider IT environment involving any node that can host and execute the corresponding program, whether it has a hypervisor or not (i.e. serverless paradigm[5]). For instance, it is possible to deploy and launch an *executable applicative program* on a simple laptop without interrupting the execution of its operating system. It is also possible to deploy an *application module* and to integrate it dynamically within an application code whose design is based on *components* (or *micro services)-oriented approach*[6].

This analysis allows (re)-defining the concept of network function (NF) (Fig. 3), which basically consists in a given processing of packets at any level of the communication stack (Application, Middleware, etc.).

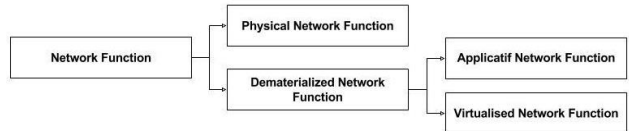

Fig. 3: Network function (NF) concept

In our vision, the concept of NF integrates and extends the concept of ETSI VNF which appears as a special case of what we call a *dematerialized network function* (DNF), i.e. deployed outside of its original environment (as opposed to the *physical network functions - PNF*). A DNF can then be defined either as a VNF or as an *application network function* (ANF) consisting in an executable applicative program or an application module.

*Software defined networking.* Initially introduced in the early 2000s, the concept of *programmable network* is now being applied by the network operators through the concept of SDN (*Software defined networking*). SDN is a framework that allows network administrators to dynamically and automatically manage a large number of network devices, services, topologies, paths, or QoS policies using an API and / or high-level languages. The targeted management includes provisioning, activation, monitoring, optimization and management of FCAPS (Fault, Configuration, Accounting, Performance, Security) in a multi-tenant network environment.

*General problem statement.* In this new ecosystem, several challenges already addressed in the classical Internet are to be (re)-considered, in particular the performance capability of the communication system in response to the expressed application-level QoS requirements. The whole complexity comes from several considerations:
- the QoS issue has to be taken into account at several levels of the communication stack, typically network-oriented levels and middleware level; the consistency of adaptation choices is an issue in itself;
- QoS-oriented DNF may be deployed either inside or outside the data path. In this latter case, a specific issue comes from the need to redirect the traffic appropriately, without interruption of application;
- the IT deployment environment is heterogeneous on various points including the DNF deployment capabilities (VNF, ANF, none), and the SDN capabilities of the traversed networks; the issue is then to discover those capabilities with the aim to take benefits of them;
- both the applications QoS-requirements and the resources/capabilities of the IT deployment environment may evolve during the application execution; the issue is then to adapt the current solutions in response to the evolution of the requirements and/or the resources/capabilities.

Next Fig. 4 provides a global illustration of the targeted IT deployment environment. This one includes not only dedicated middleware and network entities but also private and public data centers on which it is possible to deploy VNF, and simple machines on which it is possible to deploy ANF. It should be noted that those NF can be applied at the Application level (in the OSI sense of the term), but also at the Transport level (i.e. TCP level of the Internet stack) if the latter is built according to a component-based design[7].

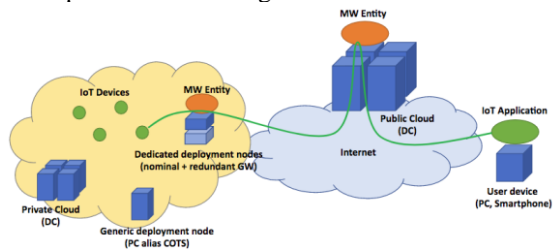

Fig. 4: Targeted deployment environment

The following paper is structured as follows. Section 2 provides a brief state of the art. Section 3 describes our general solution approach as a first contribution. One of the specific issues is then stated and the proposed solution is introduced as a second contribution. Section 4 details this specific solution. Section 5 provides the associated implementation and the experimental results. Section 6 concludes and provides our current prospective work.

## 2. Related work

Eclipse OM2M is an open source implementation of a middleware layer that is conformed to the oneM2M specification[8]. Developed at LAAS-CNRS, OM2M provides a RESTful API for XML data exchange through unreliable connections within a highly distributed environment (Fig. 5). It offers a modular architecture running on top of an OSGi Equinox runtime. OM2M provides a flexible service capability layer (SCL) that can be deployed in an M2M network, a gateway, or a device. An SCL is composed of small tightly coupled plugins, each one offering specific functionalities. A plugin can be remotely installed, started, stopped, updated, and uninstalled without requiring a reboot. It can also detect the addition or the removal of services via the service registry and adapts accordingly facilitating the SCL extension.

The CORE is the main plugin that should be deployed in each SC element. It provides a protocol-independent service for handling REST requests. Specific communication mapping plugins can be added to support multiple protocol bindings such as HTTP and CoAP. Eclipse OM2M can be extended with specific device management mapping plugins to perform device firmware updates by reusing existing protocols such as OMA-DM and BBF TR-069. It can be also extended by various interworking proxy plugins to enable seamless communication with legacy devices, such as Zigbee and Bluetooth technologies. The TLS-PSK protocol is used to secure M2M communications based on pre-shared keys. A new plugin based on the autonomic computing paradigm is designed to enhance Eclipse OM2M resource discovery and self-configuration.

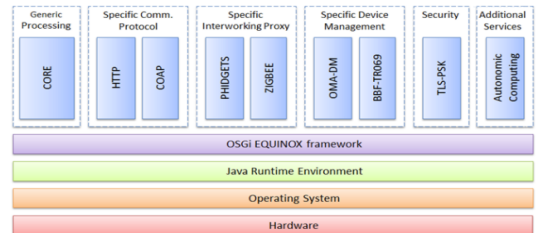

Fig. 5: Eclipse OM2M building blocks

Several other specific (i.e. not standard) MW solutions have also been proposed for which the QoS issue has been addressed. The paper[9] proposes to enhance the MW WuKong[10] for the QoS management. It introduces the concept of quality score that considers multiple QoS metrics (response time, reliability, etc.). Adequate physical devices are selected as well as their optimal deployment is decided in order to achieve the highest quality score. The limit of this approach lies in the fact that applications' QoS requirements are not taken into account dynamically. In the MiLAN project[11], Heinzelman proposes a MW that manages both nodes and the network, depending on the application description and its expressed QoS requirements. Other solutions such as[12, 13] rely on the integration of the MQTT protocol[14] for QoS management. Let us also note that the last specification of the oneM2M standard[15] proposes to integrate the MQTT protocol.

## 3. General solution approach, specific problem statement and proposed contribution

### 3.1. General solution approach

Our general solution approach consists in designing, developing and testing generic architectures for *self-adaptive* management of QoS-oriented PNF/DNF at the different levels of the communication stack:
- taking advantage of the technological opportunities associated with the dynamic deployment of PNF / DNF, but also of SDN-based networks,
- taking into account the factual heterogeneity of the solutions being deployed,
- ensuring the consistency of the (re)-configuration choices for each level through appropriate theoretical tools.

In our work, the *self-adaptive* management is based on the *autonomic computing* (AC) model[16]. This model is illustrated on the left part of Fig. 6 via the autonomic manager (AM). The AM interacts with the managed system thanks to *sensors* and *effectors*. Right part of Fig. 6 illustrates what is called the "MAPE-K loop" executed by the AM with the aim to *Monitor* the system and to raise symptoms when the expected QoS is no more reached, to *Analyze* those symptoms and to decide if a reconfiguration has to be considered, to *Plan* the adequate reconfiguration actions of the system, and finally to *Execute* those actions by the managed system, all this process being helped by a *Knowledge base* including (among other) the necessary values / rules / models/ etc. required by the MAPE algorithms.

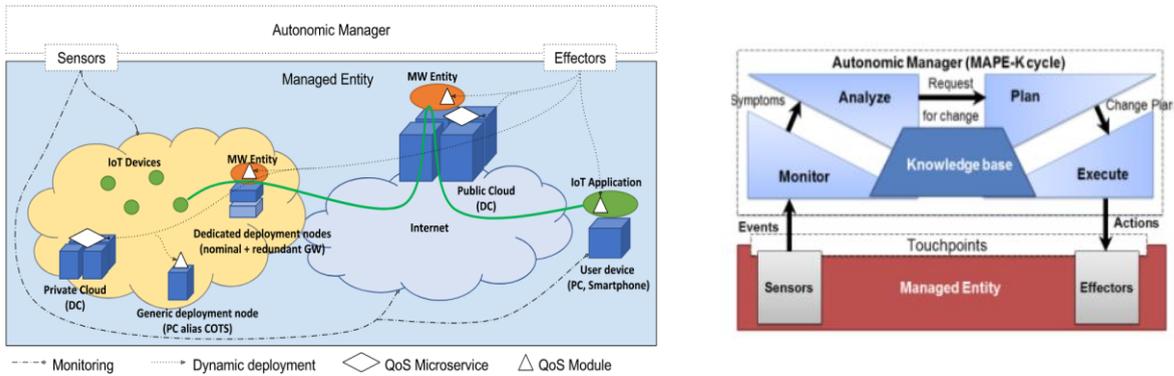

Fig. 6: Global architectural vision for a self-adaptive QoS-oriented IoT system

*3.2. Specific problem addressed in this paper and proposed contribution*

In this paper, we focus on the Middleware level. We consider a network environment involving both SDN-based networks and *legacy* (i.e. not SDN based) networks. The Middleware layer that is considered is the OM2M open source implementation presented in the related work section of this paper. We also consider that middleware level QoS-oriented mechanisms may be deployed as DNF. Some of them may then be executed outside of the data path traversed by the application level traffic. Previous Fig. 2 illustrates such a case where a (for instance) dropping/shaping/scheduling NF is supposed to be dynamically deployed on a proxy node initially outside of the data path. The issue is then to make the traffic pass within the proxy without interrupting the communication and ideally in a transparent way for the involved Server and Gateway. Conceptually, within an SDN-based network, such an issue could be easily solved thanks to an adequate dynamic redirection of the traffic by the underlying SDN switches. However, such a possibility is not provided in a legacy network.

In this context, our solution to face this specific problem is to deploy, within the server of Fig. 2, a specific ANF, whose role is to redirect the traffic without interruption of the application. Basically, the proposed solution is based on the *adapter pattern* that can be used in any OSGI implementation[17]. The following sections 4 and 5 respectively detail the proposed solution through a case study, and then show the benefits that can be induced by the performed implementation.

## 4. Design of a redirection ANF and integration within an oneM2M middleware platform

The contributions proposed in this paper are presented and assessed through the following use case. This section is organized in three subsections. In the first one, we present the considered use case. In the second one, we present the different related implementations. The third subsection presents the experimentation and the obtained results.

*4.1. Problem description through a case study*

In the field of vehicular transportation, software applications often communicate with devices (i.e. sensors and actuators) through mobile gateways (i.e. cellular networks; 3/4G). In our use case, a navigation application is transmitting 3D maps to a moving car's system (Fig. 7). In this figure, we can identify multiple elements:

- Autonomic manager: entity that implementing the control loop MAPE-K loop and allowing to decide when, where and how to deploy the implemented modules in order to meet the QoS requirements;
- Sensor: logical component allowing to retrieve the round-trip time of the IoT Application;
- Effector: logical component allowing to add/remove/modify modules into the MW nodes and the Server;
- Server/Gateway: entity that represents an infrastructure/middle node as specified in the oneM2M standard;
- Data Path: path followed by the data from the IoT application to the vehicle on-board device.

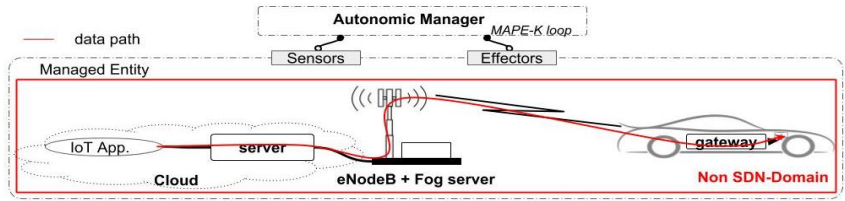

Fig. 7: Architecture of the considered use case

In this use case, vehicles are connected through cellular mobile networks and the networks that interconnects the applications and the IoT devices are supposed to be non SDN-compliant. The vehicles mobility and/or the radio conditions often lead to a significant decrease in the radio link quality, and thus a decrease of the offered QoS. The implementation of our approach for a dynamic, autonomous and seamless deployment of QoS mechanisms requires the execution of a redirection mechanism in order to put other QoS mechanisms within the data path. For example, in the NFV standard, this redirection problem is supposed to be entirely tackled by the programmability of the underlying network[18]. Indeed, in an SDN-based network, the redirection mechanism is implemented within the SDN switches and may be triggered by reconfiguration commands (i.e. OpenFlow rules)[19]. However, in a non programmable network (or in an SDN-network where reconfiguration is not available/allowed), this necessary redirection mechanism may be set up at the application layer through internal components of the targeted node. The following section presents our redirection mechanism that can be dynamically deployed within a oneM2M MW node. It can be configured to override the data path in a seamless way for the communicating applications.

### 4.2. Proposed mechanism: a generic ANF-based redirection mechanism

The redirection mechanism proposed in this paper allows re-routing the data traffic from a given entity towards another. This mechanism modifies the traffic in order to be accepted by the targeted entity. The logic behind this mechanism is presented in the following behavior diagram (Fig. 8).

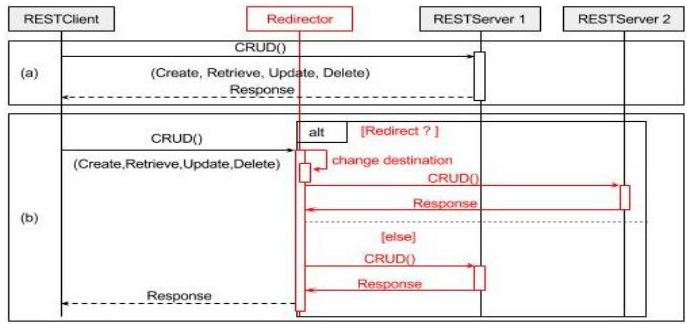
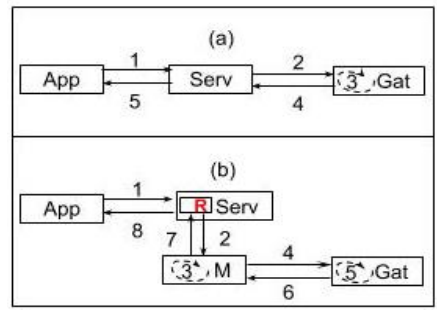

Fig. 8: Requesting a REST resource (a) through redirection mechanism (b)

Fig. 9: Implementation example of a redirection mechanism within an IoT middleware

The proposed mechanism allows redirecting requests/responses for REST resources. For example, in a communication between an application and a gateway going through a server (Fig. 9 - part a), this mechanism is deployed as a software module within the server in order to redirect the traffic toward an entity M (Fig. 9 - part b).

### 4.3. Integration to a oneM2M middleware

The oneM2M functional architecture identifies logical entities dubbed MW nodes (typically server /gateways of previous Fig. 1), each one offering a portion of the MW service. It is a resource-oriented architecture where the functionalities of the system are exposed by means of APIs. Each of these entities is composed of software modules that implement each one of the node's features. Thus, based on this modular architecture of the MW node, we propose to integrate the new mechanisms as modules. These modules can be incorporated dynamically at design or run time in a seamless fashion without any modification of the original MW node. Fig. 10 presents a simplified view of the architecture of a oneM2M middleware (here Eclipse OM2M) node having the redirection mechanism. More details about this architecture can be found in[6]. In the following section, we apply our approach to the Eclipse OM2M middleware platform.

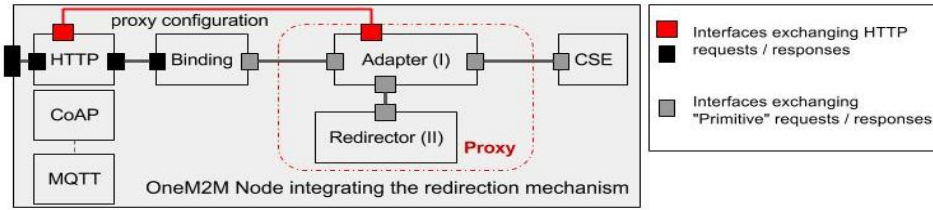

Fig. 10: Integration architecture of a redirection mechanism (simplified view)

## 5. Implementation and performance evaluation

### 5.1 Implementation

This section presents the implemented algorithms of the different ANFs. These ANFs have been implemented in the form of software modules following the OSGi standard.

*5.1.1. Implementation of the Redirection ANF*

| Algorithm 1: Redirector | Description |
|---|---|
| **Input**: message; pattern.<br>**Output**: message.<br>0: **begin**<br>1: **Initialize** p ←message.destination;<br>2: **if** (p ≠ null) **then**<br>3:     **if** (p is a key of pattern) **then**<br>4:         message.destination ← pattern[p]<br>5:         response ← Send message to pattern[p]<br>6:     **end if**<br>7: **end if**<br>8: **return** response<br>9: **end** | When the redirection ANF receives a message:<br>● It retrieves its destination (line 1),<br>● Based on a match, it chooses the traffic to redirect (line 3). This matching is achieved based on the destination resource of the message and a policy that is provided by a third-party entity.<br>● It changes the destination of the message according to this policy (line 4)<br>● It redirect the message towards the entity indicated in the policy (line 5), then returns the response to the message sender. |

*5.1.2. Implementation of the compression/decompression ANF*

| Algorithm 2: Compressor | Description |
|---|---|
| **Input**: message; destination.<br>**Output**: message.<br>0: **begin**<br>1: **Initialize** p ←message.payload;<br>2: message.payload ← *DEFLATE*(p)<br>3: response ← Send message to destination<br>4: **return** response<br>5: **end** | The compression ANF achieves three tasks:<br>● It receives the message and extracts its payload (line 1)<br>● It compresses the payload following a lossless data compression algorithm (i.e. "DEFLATE"[20]) (line 2)<br>● It builds the message and transfers it. (line 3) |
| **Algorithm 3: Decompressor** | **Description** |
| **Input**: message.<br>**Output**: message.<br>0: **begin**<br>1: **Initialize** p ←message.payload;<br>2: message.payload ← *INFLATE*(p)<br>3: **return** message<br>4: **end** | The decompression ANF achieves three tasks:<br>● It receives the message and extracts its payload (line 1)<br>● It decompresses the payload following the "INFLATE" algorithm [20] (line 2),<br>● It builds the message and transfers it (line 3) |

## 5.2 Experimentations and performances evaluation

The objective of the experimentations being presented in this section is to validate our redirection mechanism and to evaluate the benefits of such mechanisms along other mechanisms such as data compression. To this end, we have evaluated the round-trip time between the initial communicating entities.

### 5.2.1. Scenario description

The considered scenario takes place in several steps:
- 1) the (mobile) vehicle is located in a geographical zone where mobile network coverage is optimal;
- 2) the AM predicts a QoS degradation (represented here as a value of RTT > 1s) of an IoT application interacting with this vehicle and reacts by dynamically deploying a traffic compression ANF (**C**) within a hosting Fog node (ANF compliant) near the cellular network base station. The AM jointly deploys a traffic redirection ANF (**R**) from the IoT Server towards the Fog node, and a decompression ANF (**U**) within the IoT Gateway, without any service interruption and in a seamless way for the application (Fig. 11);
- 3) the vehicle gets away from the base station, leading to a degradation of the IoT application QoS.

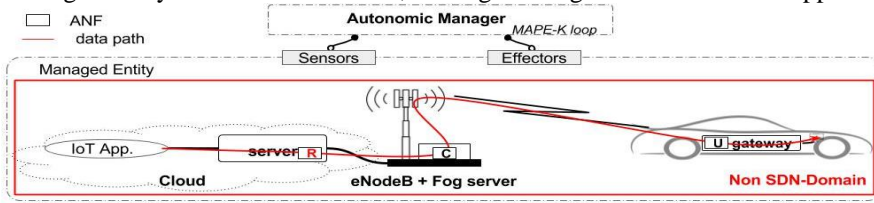

Fig. 11: Second step of the scenario

### 5.2.2. Results and discussion

Next Fig. 12 presents the evolution of the available bandwidth between the base station and the vehicle. It also presents the evolution of the IoT application response times. Every point is the result of an average of measures taken during 20 seconds. We can observe a comparison of average throughputs obtained by the application without adaptation (RTT_1), then with adaptation (RTT_2).

On the figure, we can clearly observe two stages:
- stage 1: application response times are stable and constant (timestamp #1 to #6 for RTT_1 metric, and timestamp #1 to #11 for RTT_2 metric). During this stage, the obtained response times without adaptation are higher (~ 0,5s) than those obtained with adaptation (~ 0,2s);
- stage 2: response times increase with the drop of the available bandwidth (timestamp #7 to #15 for RTT_1 metric and timestamp #11 to #13 for RTT_2 metric. During this stage, the compressed and redirected traffic allows to obtain response times that are less than 1s (i.e. the tolerated threshold by the IoT application). Without this adaption, response times exceed 2s.

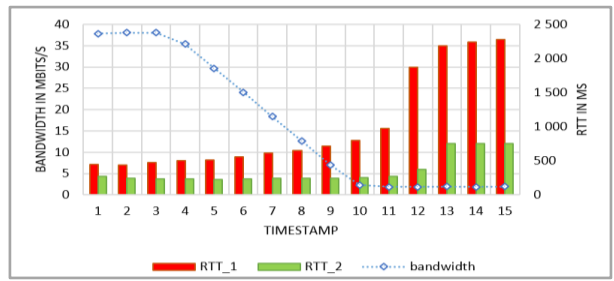

Fig. 12. Average response time obtained by the IoT application with and without adaptation

## 6. Conclusions and future works

QoS-oriented considerations will have to be tackled at different levels of the communication stack for the future middleware IoT applications. In parallel, the advent of virtualization and softwarization technologies open new opportunities to tackle this issue. Deployment of network functions (NF) is now to be really considered. In the same way, network operators will progressively "open" their network, allowing their users to "program" the network depending on their requirements. As a result, the near future Internet will be even more heterogeneous than the current one, both in terms of NF deployment solution and in terms of network programing capabilities. In this context, this paper first detailed this general problematic. It then presented our vision for a dynamic and autonomic management of the QoS required by IoT applications, distributed in such heterogeneous environment. Finally, we exposed a case study illustrating how to tackle a new problem dealing with the implementation of a QoS-oriented NF "outside" of the data path followed by the traffic generated by a middleware-based application. We designed a redirection NF that is dynamically plugged without interrupting the data transfer. We also showed through experimental results the benefits that can be induced by such an adaptation. Our current and future work deal with several points related to the design of the architecture of the targeted autonomous system, and the elaboration of theoretical models to drive the monitoring, the analysis and the planning phases of the MAPE-K loop.